\documentclass[11pt,a4paper,english,nofootinbib,,superscriptaddress]{revtex4}
\usepackage{lmodern}

\usepackage[T1]{fontenc}
\usepackage[latin9]{inputenc}
\usepackage{babel}

\usepackage{amsmath}
\usepackage{graphicx}
\usepackage{amssymb}
\usepackage{esint}
\usepackage[unicode=true, pdfusetitle,
 bookmarks=true,bookmarksnumbered=false,bookmarksopen=false,
 breaklinks=false,pdfborder={0 0 1},backref=false,colorlinks=false]
 {hyperref}
\usepackage{amsmath}
\usepackage{amssymb}
\def\b{\begin{equation}}
\def\e{\end{equation}}

\def\N{\mbox{I\hspace{-.15em}N}}

\def\Z{\mbox{Z\hspace{-.3em}Z}}

\def\R{{\rm I\hspace{-.15em}R}}

\def\C{\hspace{3pt}{\rm l\hspace{-.47em}C}}

\def\b{\begin{equation}}
\def\e{\end{equation}}
\def\bee{\begin{enumerate}}
\def\eee{\end{enumerate}}
\def\1{\mbox{1\hspace{-.25em}I}}

\def\N{\mbox{I\hspace{-.15em}N}}

\def\Z{\mbox{Z\hspace{-.3em}Z}}

\def\R{{\rm I\hspace{-.15em}R}}

\def\C{\hspace{3pt}{\rm l\hspace{-.47em}C}}

\def\bd{\begin{displaystyle}}
\def\ed{\end{displaystyle}}
\def\ba{\begin{array}}
\def\ea{\end{array}}

\def\bee{\begin{enumerate}}
\def\eee{\end{enumerate}}

\def\bes{\begin{eqnarray*}}
\def\ees{\end{eqnarray*}}
\def\be{\begin{eqnarray}}
\def\ee{\end{eqnarray}}

\makeatletter
\@ifundefined{textcolor}{}
{%
 \definecolor{BLACK}{gray}{0}
 \definecolor{WHITE}{gray}{1}
 \definecolor{RED}{rgb}{1,0,0}
 \definecolor{GREEN}{rgb}{0,1,0}
 \definecolor{BLUE}{rgb}{0,0,1}
 \definecolor{CYAN}{cmyk}{1,0,0,0}
 \definecolor{MAGENTA}{cmyk}{0,1,0,0}
 \definecolor{YELLOW}{cmyk}{0,0,1,0}
 }

\usepackage{latexsym}\usepackage{bm}

\makeatother

\begin{document}

\title{Entropy of Quantum Fields \\ in de Sitter Space-time}

\author{M.V. Takook}

\email{mtakook@yahoo.com; takook@razi.ac.ir}
 
\affiliation{Department of Physics, Razi University,
Kermanshah, IRAN}

\date{\today}

\begin{abstract}

The quantum states or Hilbert spaces for the quantum field theory in de Sitter space-time are studied on ambient space formalism. In this formalism, the quantum states are only depended $(1)$ on the topological character of the de Sitter space-time, {\it i.e.} $\R \times S^3$, and $(2)$ on the homogeneous spaces which are used for construction of the unitary irreducible representation of de Sitter group. A compact homogeneous space is chosen in this paper. The unique feature of this homogeneous space is that its total number of quantum states, ${\cal N}$, is finite although the Hilbert space has infinite dimensions. It is shown that ${\cal N}$ is a continuous function of the Hubble constant $H$ and the eigenvalue of the Casimir operators of de Sitter group. The entropy of the quantum fields on this Hilbert space have been calculated which is finite and invariant for all inertial observers on the de Sitter hyperboloid. 
\end{abstract}
\maketitle

\vspace{0.5cm}
{\it Proposed PACS numbers}: 04.62.+v, 03.70+k, 11.10.Cd, 98.80.H

\tableofcontents

\section{Introduction}

An inertial observer in the de Sitter (dS) vacuum state
detects a thermal radiation at the temperature $ T=\frac{H}{2 \pi}$ for each coordinate system on the dS hyperboloid. Its corresponding thermodynamic entropy is
\cite{giha} \b \label{dse}
S_{dS}=\frac{A_c}{4G}=\frac{\pi}{GH^2}=\frac{1}{4\pi G T^2},\e where
$G$ is Newton's constant and $A_c=4\pi H^{-2}$ is the area of the cosmological event horizon in dS space. This entropy is similar to the
Bekenstein-Hawking entropy of the black hole: $ S_{BH}=\frac{ k_B c^3}{ \hbar}\frac{ A_h}{4G}, $ where $A_h$ is the area of black hole horizon, $c$ is the speed of light, $k_B$ is the Boltzmann constant and $\hbar=\frac{h}{2\pi}$
is the Planck constant. The microscopic origin of this entropy is an outstanding problem in theoretical physics \cite{stva,wit,spstvo,novawe}. The structure of the Hilbert space of the physical system, which includes the gravitational field is unknown mostly because of the lack of a proper method to quantize the gravity. Also, the Hilbert space in quantum field theory of curved space-time has infinite dimension. Then the total number of the quantum states is infinite and a finite entropy cannot be obtained. These problems have been surveyed in different methods by several authors \cite{bomast,mari,ardunitr,wit,spstvo,mast,bafi,bafipa,pave,pave2,kakalitr,
novawe,nom,joshsi,dav,mipa}.

In this paper, the microscopic origin of the dS entropy, in the context of quantum field theory (QFT) in dS ambient space formalism is considered and the total number of quantum states has been calculated. For this purpose we must explicitly construct the Hilbert spaces for QFT in dS ambient space formalism. We have studied the QFT in dS ambient space formalism, which allows one to consider the QFT in a rigorous mathematical framework, based on the analyticity of the complexified pseudo-Riemannian manifold and the group representation theory. In this formalism some interesting results are obtained. Using the charge conjugation symmetry in the dS ambient space notation a novel de Sitter super-symmetry algebra in the ambient space notation is obtained  \cite{morrota,parota}. We can simply generalize the gauge theory, super-symmetry and super-gravity in the dS ambient space notation in a direct way from to the Minkowskian counterpart \cite{ilio,frva,ta14,rota05}. 

As the usual, the quantum fields are divided into two distinguishable parts: ''massive'' and ''massless'' fields noting that the concept of mass is not well defined in dS space because mass is the eigenvalue of the Casimir operator of the Poincar\'e group. Nevertheless, the principle of causality is well defined in the dS space and elegantly in the
ambient space formalism \cite{brgamo,brmo,ta97}. Here a field is called massive, when it propagates inside the dS light-cone and corresponds to the massive Poincar\'e field in the null curvature limit. On the other hand, when it propagates on the dS light-cone it is called massless and corresponds to the massless Poincar\'e field at $H=0$. 

The dS group $SO(1,4)$ has three different types of representations: principal, discrete and complementary series representations \cite{tho,new,dix,tak,lip,vikl}. It has been shown that the massive various spin fields ($s=0,\frac{1}{2}, 1, \frac{3}{2}, 2$) in the dS ambient space formalism are associated with the principal series representation of the dS group \cite{gata,ta96,bagamota,gagata,taazba}. The massless conformally coupled scalar fields in the dS space correspond to the complementary series representation of the dS group \cite{brmo}.
The massless spinor field ($s=\frac{1}{2}$) corresponds to the lowest representation of the discrete series \cite{ta97,bagamota} and it is also conformal invariant. The massless fields with $s= 1, \frac{3}{2}$ in the dS space beside conformal invariance are also gauge invariant and they correspond to the indecomposable representation of the dS group but their physical states (or central parts) are correspond to the lowest representation of the discrete series  \cite{gagarota,berotata,azam,fatata}.

The massless spin-2 field in the dS space (linear
quantum gravity) is of great importance since it has an essential role in the quantum cosmology and the quantum gravity on the dS space. It has been found that its corresponding propagator in the linear approximation exhibits a pathological behaviour for large separated points (infrared divergence). Antoniadis, Iliopoulos and Tomaras have shown that
the large-distance pathological behavior of the graviton propagator on the dS background does not manifest itself in the quadratic part of the effective action in the one-loop approximation \cite{anilto}. This means that the pathological behaviour of the graviton propagator may be gauge dependent and should not appear in an effective way as a physical quantity. In dS ambient space notation, the massless spin-2 field can be written in terms of a minimally coupled scalar field \cite{ta97,taro12}. The massless minimally coupled scalar fields in the dS space correspond to the indecomposable representation of the dS group \cite{ta14,gareta}. The Krein space quantization (Hilbert space $\oplus$ anti-Hilbert space) was presented for obtaining a covariant quantization of this field, which is free of any infrared and ultraviolet divergences \cite{bognar,min,azio,dere,gareta,ta01}.
Then the linear gravity or massless spin-$2$ field, which are described by a rank-$2$ symmetric tensor field $K_{\alpha\beta}$ in ambient space formalism can be quantized on Gupta-Bleuler vacum state (Krein space quantization) \cite{ta99,derotata,rota04,rota,ta09,taro12,petata} or on Bunch-Davies vacuum state without any infrared divergences \cite{ta14,taro14}. 

The massless field, which propagates on the dS light-cone must be conformal invariant. Such a field is supposed to transform simultaneously under the unitary irreducible representation (UIR) of the dS group and the conformal group $SO(2,4)$. The rank-$2$ symmetric tensor field $K_{\alpha\beta}$ (linear gravity) in the ambient space notation (or in the Dirac's $6$-cone formalism) cannot be transformed simultaneously under the UIR of the dS and the conformal groups \cite{bifrhe,tatafa,tata}. The linear gravity (or the conformal gauge gravity in the dS background), which transforms simultaneously under the UIR of the dS and the conformal groups, is a spin-$2$ rank-$3$ mixed symmetry tensor field ${\cal K}_{\alpha\beta\gamma}$ \cite{tatafa,tata,tapeta,ta14}. This field is also gauge invariant and then it corresponds to the indecomposable representation of the dS group, though its physical states (or central parts) correspond to the lowest representation of  the discrete series of the dS and the conformal groups \cite{ta14}. Then the gravitational field is divided into two different parts. The first part is the dS metric or the dS background ($g_{\mu\nu}^{dS}$). This part behaves classically and forms the structure of the space-time as a base manifold. The effect of this part on the quantum field appears as a UIR of the dS group. The remaining part is a spin-$2$ rank-$3$ mixed symmetry tensor field ${\cal K}_{\alpha\beta\gamma}$. This part can be considered as a conformal gauge gravity on the dS background which is a fibre principal with a gauge group $SO(2,4)$ on the base dS manifold \cite{ta14,hel,chchde,nak}. The propagator of this field is free of any infrared divergence. It is interesting to note that the various-spin massless fields in Dirac $6$-cone formalism can be simply mapped to the massless fields in the dS hyperboloid in the ambient space formalism. We show that the homogeneous space of the UIR of the discrete series of the conformal group and the dS group, is the same. This homogeneous space provides the best basis for the Hilbert space of discrete series representation, which is called Graev's realization of the discrete series.

In the previous papers, we have constructed the quantum field operators using the Wightman two-point function in the ambient space formalism. Here we build directly the Hilbert space in the dS ambient space notation from the UIR of the dS group. The quantum states are only depend on the homogeneous space where the UIR of the dS group has been constructed on it as well as the topological character of dS space {\it i.e.} $\R \times S^3$. There are different realizations for the construction of the UIR of the dS group and their corresponding Hilbert spaces. There exist the homogeneous spaces with the finite total volume. Each point in these homogeneous spaces represents a vector in the Hilbert spaces.

Since a maximum length for an observable or an even horizon in dS space exist then a minimum size in the compact Homogeneous space can be defined by using the Heisenberg uncertainty principle. Each point in this space represents a vector in Hilbert space and the number of points is infinite mathematically. Since the total volume of Homogeneous space is finite and a minimum length in this space exists from uncertainty principle, therefore the total number of points become finite physically. It means that the total number of quantum states in these Hilbert spaces is finite. The total number of the quantum states, ${\cal N}$, is the sum of (or the integration over) the admissible value of the points in the homogeneous space and for the compact homogeneous space it is finite, although the Hilbert space has infinite dimensions. 

The total number of quantum states, ${\cal N}$, is a function of the Hubble parameter and the eigenvalues of the Casimir operators of the dS group. Since the Hilbert space is constructed on a homogeneous space, which is invariant under the scale transformation, then an arbitrary scale appears in the total number of quantum state ${\cal N}$. This arbitrariness can be fixed by imposing some physical conditions such as the Holographic principle or the entropy bounds \cite{th,su,suli,str,ngst,sre,noobod,marara,hast,alkasi,alal,bou,bou2}. Finally, the entropy of the quantum fields is calculated, which is finite. It is a function of the Hubble parameter and the eigenvalues of the Casimir operators of the dS group. These parameters are the dS invariant and equivalent for each inertial observer.

The organization of this paper is as
follows: In the next section, we present the notations including two independent Casimir operators of the dS group. Section III is devoted to a brief review of the partial wave solution for the massive scalar field in the dS intrinsic coordinates. In addition, the Hilbert space $({\cal H}_L)$ has been presented and the total number of quantum states ${\cal N}_L$ (which is infinite) has been calculated. The solution of the massive scalar field equation in the dS ambient space formalism (plane waves) is presented in section IV and it has been shown that this
solution can be written in terms of a constant five-vector $\xi^\alpha$. This five-vector is a homogeneous space where the UIR of the dS group or the Hilbert space is constructed on it. We demonstrate that there exist a specific realization for $\xi^\alpha$ and then for the basis of the Hilbert space which includes a finite total number of quantum states. In section V, the relation between the partial waves and the dS plane waves is presented. We generalize the construction of the Hilbert space in the ambient space notation to the various spin fields in section VI. This construction is based on the UIR of the de Sitter group. Section VII is allocated to a brief review of the Dirac $6$-cone formalism and the discrete series representation of the conformal group $SO(2,4)$ for the massless field. In section VIII, the entropy of a massive quantum fields is calculated and the possibility of fixing the arbitrariness by the holographic principle or entropy bounds and two other physical conditions have been discussed. Finally a brief conclusion and an outlook for further
investigation have been presented.

\setcounter{equation}{0}
\section{Notation}

In this section we briefly introduce the notation and convention, which are being used in this paper. The dS space-time can be identified by a 4-dimensional hyperboloid embedded in 5-dimensional Minkowskian space-time with the constraint:
     \b M_H=\{x \in \R^5 ;x^2=\eta_{\alpha\beta} x^\alpha
 x^\beta =-H^{-2}\},\;\; \alpha,\beta=0,1,2,3,4, \e
where $\eta_{\alpha\beta}=$diag$(1,-1,-1,-1,-1)$. The dS metrics is \b  ds^2=\eta_{\alpha\beta}dx^{\alpha}dx^{\beta}|_{x^2=-H^{-2}}=
g_{\mu\nu}^{dS}dX^{\mu}dX^{\nu},\;\; \mu=0,1,2,3,\e where $X^\mu$
are 4 space-time intrinsic coordinates system on dS hyperboloid. Different coordinate systems can be defined in the dS space. In this paper, we chose the bounded global conformal coordinates $(S^1 \times S^3)$, which is defined by: \b \label{csc} \left\{ \ba{rcl}
x^0&=&H^{-1}\tan \rho\\
x^1&=&(H\cos\rho)^{-1}\,(\sin\alpha\,\sin\theta\,\cos\phi),\\
x^2&=&(H\cos\rho)^{-1}\,(\sin\alpha\,\sin\theta\,\sin\phi),\\
x^3&=&(H\cos\rho)^{-1}\,(\sin\alpha\,\cos\theta),\\
x^4&=&(H\cos\rho)^{-1}\,(\cos\alpha),\ea\right.\e where
$-\pi/2<\rho<\pi/2$, $0\leq\alpha\leq\pi$, $0\leq\theta\leq\pi$
and $0\leq\phi \leq 2\pi$ and $X^\mu\equiv (\rho, \alpha, \theta, \phi),\;\mu=0,1,2,3$. The dS metrics now can be written as \b
ds^2=\frac{1}{H^2 \cos^2\rho}
(d\rho^2-d\alpha^2-\sin^2\alpha\,
d\theta^2-\sin^2\alpha\sin^2\theta \,d\phi^2)=\frac{1}{H^2 \cos^2\rho}
(d\rho^2-d\Omega_{S^3}),\e
where $d\Omega_{S^3}$ is the metric on three-sphere $S^3$. The real de Sitter global coordinate is $(R \times S^3)$: $x^\alpha=(H^{-1}\sinh Ht, H^{-1}\cosh Ht\;\; {\bf u})$, where $(u^1)^2+(u^2)^2+(u^3)^2+(u^4)^2=1$. For this coordinate system the metric is \cite{mot}: 
\b \label{gcor} ds^2=dt^2-(H^{-1}\cosh Ht)^2 d\Omega_{S^3},\e
where the spatial sections is a three-sphere with radius $H^{-1}\cosh Ht$.

Any geometrical object in dS space can be written either in terms of the four local coordinates $X^\mu$ (intrinsic coordinates)
or the five global coordinates $x^\alpha$ (ambient space formalism) with the two supplementary conditions: homogeneity and transversality. The tensor or spinor fields in the dS space are the homogeneous function in the ambient space formalism \cite{dir}
\b \label{homog} \Phi^i_{\alpha_1,...\alpha_n}(\lambda x)=\lambda^\sigma \Phi^i_{\alpha_1,...\alpha_n}(x),\e
where $i$ is the spinorial index ($i=1,2,3,4)$, and $\alpha_l$ is the tensorial index. The $\Phi$ is said to be a homogeneous function of degree $\sigma$. The transversality condition is \cite{dir}: $$ x^{\alpha_1}\Phi^i_{\alpha_1,...\alpha_n}(x)=0.$$

The action of the dS group $SO(1,4)$ on the intrinsic coordinates $X^\mu=(\rho,\alpha,\theta,\phi)$ is utterly complicated but on the ambient space coordinates $x^\alpha$ is rather simple:
$$ x'^\alpha=\Lambda^\alpha_\beta x^\beta, \;\;\;\Lambda \in SO(1,4) \Longrightarrow x.x=x'.x'=-H^{-2}, $$ where $ SO(1,4)=\left\lbrace \Lambda \in GL(5,\R)| \;\; \det \Lambda=1,\;\; \Lambda \eta \Lambda^t= \eta \right\rbrace$. The ambient space coordinate $x^\alpha$ can be define by a $4\times 4$ matrix $X$:
$$ X=\left( \begin{array}{clcr} x^0 & {\bf p} \\ {\bf \bar p} & x^0 \\    \end{array} \right),$$ where ${\bf p}\equiv(x^4, \vec x)$ is a quaternion and ${\bf \bar p}\equiv(x^4,-\vec x)$ is its quaternion conjugate \cite{tak,ta97}:
\b \label{quat} {\bf p}=(x^4, \vec x)\equiv\left( \begin{array}{clcr} x^4+ix^3 & ix^1-x^2 \\ ix^1+x^2 & x^4-ix^3 \\    \end{array}, \right), \;\; {\bf \bar p}={\bf p}^\dag.\e  The quaternion norm is  $|{\bf p}|=({\bf p}{\bf \bar p})^{\frac{1}{2}}=\sqrt{(x^1)^2+(x^2)^2+(x^3)^2+(x^4)^2}$. If we defined ${\bf p}=H^{-1}\cosh Ht \;\;{\bf u}$, we obtain the metric (\ref{gcor}) and ${\bf u}$ is a quaternion with the norm $|{\bf u}|=1$.

Matrix $X$ under the  group $Sp(2,2)$ transforms as \cite{tak}:
$$ X'=gX\bar g^t, \;\;\;g \in Sp(2,2),\;\; \det X=\det X' \Longrightarrow x.x=x'.x'=-H^{-2},$$
where the group $Sp(2,2)$ is defined by: \b \label{sp22} Sp(2,2)=\left\lbrace g=\left( \begin{array}{clcr} {\bf a} & {\bf b} \\ {\bf c} & {\bf d} \\    \end{array} \right), \;\; \det g=1 , \;\; J \bar g^t J=g^{-1}, J=\left( \begin{array}{clcr} \1 &\;\; 0 \\ 0 & -\1 \\    \end{array} \right)\right\rbrace . \e 
$\1$ is the $2 \times 2$ matrix. The elements ${\bf a},{\bf b},{\bf c}$ and ${\bf d}$ are quaternion and $Sp(2,2)$ is the universal covering group of $SO(1,4)$:
 \b SO_0(1,4)  \approx Sp(2,2)/ \Z_2 .\e 
 For $4 \times 4$ matrix $g$, we have $\bar g^t=g^\dag$. By using the definition  (\ref{sp22}), we obtained
 $$ g^{-1}=\left( \begin{array}{clcr} \;{\bf \bar a} & -{\bf \bar c} \\ -{\bf \bar b} &\;\;\; {\bf \bar d} \\    \end{array} \right), \;\; {\bf \bar a}{\bf b}={\bf \bar c}{\bf d}, \;\; |{\bf a}|^2-|{\bf c}|^2=1,\;\; |{\bf d}|^2-|{\bf b}|^2=1. $$  The ambient space coordinates $x^\alpha$ under the action of the  group $Sp(2,2)$ transforms as \cite{tak}:
 \b \label{sp2tra} \left\lbrace \begin{array}{clcr} x_0'=(|{\bf a}|^2+|{\bf b}|^2)x_0 & +{\bf a}{\bf p}{\bf \bar b}+{\bf b}{\bf \bar p}{\bf \bar a}, \\ {\bf p}'=({\bf a}{\bf \bar c}+{\bf b}{\bf \bar d})x_0+ & {\bf b}{\bf \bar p}{\bf \bar c}+{\bf a}{\bf p}{\bf \bar d}. \\    \end{array} \right. \e
 
Here two homogeneous spaces are being used to construct the UIR of the dS group. For principal series the quaternion ${\bf u}=(u^4, \vec u)$ with the norm $|{\bf u}|=1$, is used and for discrete series the quaternion ${\bf q}=(q^4, \vec q)$ with the norm $|{\bf q}|<1$, is used. Their transformation under the group $Sp(2,2)$ are \cite{tak}:
$${\bf u}'= g.{\bf u}=({\bf a}{\bf u}+{\bf b})({\bf c}{\bf u}+{\bf d})^{-1}, \;\;\; |{\bf u}' |=1,$$ \b \label{qt} {\bf q}'=g.{\bf q}=({\bf a}{\bf q}+{\bf b})({\bf c}{\bf q}+{\bf d})^{-1},\;\;\; |{\bf q}'|<1.\e These quaternions can be considered as the parts of a null five vector $\xi^\alpha=(\xi^0, \vec \xi, \xi^4)$ ($\xi.\xi=0$), in a unique way, by the following relations:
\b \label{2orbit} \xi^\alpha_u \equiv (\xi^0, \xi^0 \; {\bf u}),\;|{\bf u}|=1;\;\;\; \xi^\alpha_B \equiv (\zeta \sinh \kappa, \zeta\cosh \kappa \; {\bf q}),\; \; |{\bf q}|=|\tanh \kappa|<1.\e Since $\xi.\xi=0$, the $\xi^0$ and $\zeta$ are completely arbitrary from the mathematical point of view {\it i.e.} they are scale invariant, then their transformations under the dS group are unimportant for us here.

The dS group has two Casimir operators
 \b  Q^{(1)}=-\frac{1}{2}L_{\alpha\beta}L^{\alpha\beta},\;\;  \alpha
               =0,1,...,4, \e
      \b Q^{(2)}=-W_\alpha W^\alpha\;\;\;,\;\;\;W_\alpha =\frac{1}{8}
      \epsilon_{\alpha\beta\gamma\delta\eta} L^{\beta\gamma}L^{\delta\eta},\e
where  $\epsilon_{\alpha\beta\gamma\delta\eta}$ is the usual anti-symmetric tensor in $\R^5$, $L_{\alpha\beta}$s are the infinitesimal generators of the dS group. These operators commute with the action of the group
generators and, as a consequence, they are constant on each UIR of dS group. The UIRs of the dS group are classified by the eigenvalues of the Casimir operators \cite{tho,new,dix,tak}:
             \b \label{qe1} Q^{(1)}_{j,p}=\left[-j(j+1)-(p+1)(p-2)\right]I_d,  \e
           \b Q^{(2)}_{j,p}=[-j(j+1)p(p-1)]I_d. \e
There exist three types of representations corresponding to the different values of the parameters $j$ and $p$:
\begin{itemize}
\item{Discrete series representation}
$$  j\geq p\geq1\;\mbox{ or}\; \frac{1}{2}, \;\;\; j-p=\mbox{integer number}, $$
     \b     j=1,2,3,...,\;\;\;\; \;\; p=0.\e
\item{Principal series representation}
 \b\label{ps}  j=0,\frac{1}{2},1,\frac{3}{2},2,....,\;p=\frac{1}{2}+i \nu,\;
         \;\;
       \left\{ \begin{array}{clcr} \nu \geq 0 \;  \;\mbox{for}\; \; j=0,1,2,....\\
  \nu > 0\; \;\mbox{for} \;\; j=\frac{1}{2},\frac{3}{2},\frac{5}{2},.... \\ \end{array} \right. .\e
\item{Complementary series representation}
$$  j=0\;,\;\;-2 <  p-p^2<\frac{1}{4},$$
          \b  j=1,2,3,...\;,\;\;\;0<p-p^2 <\frac{1}{4}.\e
\end{itemize}
In the following, first we recall the structure of the Hilbert space for massive scalar fields in the intrinsic coordinate system and ambient space formalism, then we generalize the ambient space notation for other spin fields.

\setcounter{equation}{0}
\section{Massive scalar field: partial waves}

A massive scalar field is defined through the following field equation:
 \b\label{sfe} [\Box_H +(m_H^2 + 12H^2 \varepsilon )]\Phi(X)=0,\e
where $\Box_H$ is the Laplace-Beltrami operator on dS space and $\varepsilon $ is the coupling between the gravitational and scalar fields. The solution of this field equation in the coordinate system (\ref{csc}) can be written as
\cite{chta,kiga,gareta}: \b\phi_{Llm}(X)=\chi_{L}(\rho){\cal Y}_{Llm}(\Omega),\e
where  ${\cal Y}_{Llm}(\Omega)$'s is the hyper-spherical harmonics \cite{chta,kiga}:
$${\cal Y}_{Llm}(\Omega)
=\left(\frac{(L+1)(2l+1)(L-l)!}{2\pi^2(L+l+1)!}\right)^{\frac{1}{2}}
2^ll!\left(\sin\alpha\right)^lC_{L-l}^{l+1}\left(\cos\alpha\right)
Y_{lm}(\theta,\phi),$$
for $(L,l,m)\in\N\times\N\times\Z$ with $0\leq l\leq L$ and $-l\leq
m\leq l$. In this equation the $C_n^\lambda$s are Gegenbauer
polynomials and 
$$Y_{lm}(\theta,\phi)=
(-1)^m\left(\frac{(l-m)!}{(l+m)!}\right)^{\frac{1}{2}}
P_l^m(\cos\theta)e^{im\phi},$$
where $P_l^m$s are the associated Legendre functions.
The ${\cal Y}_{Llm}$'s obey the
orthogonality conditions:
$$\int_{\rm
S^3}{\cal Y}_{Llm}^{*}(\Omega){\cal Y}_{L'l'm'}(\Omega)\,d\Omega
=\delta_{LL'}\delta_{ll'}\delta_{mm'}\ .$$
The radial-part $\chi_L(\rho)$ is
\b \chi_{ L}(\rho)=A_L(\cos
      \rho)^{\frac{3}{2}}\left[P^{\lambda}_{L+\frac{1}{2}}(\sin\rho)-
\frac{2i}{\pi}
    Q^{\lambda}_{L+\frac{1}{2}}(\sin\rho)\right],\e
where
\begin{eqnarray} \label{lambda}
\lambda&=&\sqrt{\frac{9}{4}-\kappa}\;\;\;\mbox{ when
}\;\;\;\frac{9}{4}\geq\kappa\geq0,\nonumber\\
\lambda&=&i\sqrt{\kappa-\frac{9}{4}}\;\;\;\mbox{ when
}\;\;\;\frac{9}{4}\leq\kappa, \;\; \kappa=m^2H^{-2}+12\varepsilon,\end{eqnarray}
and $A_L$ is given by
    $$ A_L=H\frac{\sqrt\pi}{2}\left(\frac{\Gamma(L-\lambda+\frac{3}{2})}
    {\Gamma(L+\lambda+\frac{3}{2})}\right)^{\frac{1}{2}}.$$
We then obtain the complete set of modes
\b \label{pmc}\phi_{Llm}^\lambda(X)=\chi_{L}(\rho){\cal
Y}_{Llm}(\Omega),\  X=(\rho,\Omega)\in M_H,\e
which verify the orthogonality prescription:
$$\left<\phi_{L'l'm'}^\lambda,\phi_{Llm}^\lambda\right>
=\delta_{LL'}\delta_{ll'}\delta_{mm'}\mbox{ and }
\left< \phi_{L'l'm'}^\lambda,\left(\phi_{Llm}^\lambda\right)^{*}\right>=0.$$

These modes can be used to define the Euclidean vacuum in the standard terminology. The quantum field operator for these modes can be written as:
$$\Phi(X)=\sum_{Llm}\left[ a_{Llm,\lambda}\phi_{Llm}^\lambda(X)+ a_{Llm,\lambda}^\dag \left(\phi_{Llm}^\lambda(X)\right)^*\right].$$
The vacuum states is defined by
$$  a_{Llm,\lambda}\left|O_L\right>=0,$$
and the ''one particle'' Hilbert space is:
$$  a_{Llm,\lambda}^\dag \left|O_L\right>=C_{Llm,\lambda} \left|Llm; \lambda \right> ,$$
where $C_{Llm,\lambda}$ is a normalization constant. This Hilbert space is called ${\cal H}_{L}$  and the identity operator in this case may be defined as:
\b \sum_{Llm}  | Llm; \lambda \left>\right< Llm;\lambda |  \equiv \1_L ,\;\; L=0,1,2,....; l=0,1,..,L; -l \leq m  \leq l.\e
From here on, we call the mode solution (\ref{pmc}) the ''partial waves''. Each fixed value of $L,l$ and $m$ represents a vector in this Hilbert space, which is infinite dimensional. The sum over all the possible values of these parameters is defined as the total number of the quantum states ${\cal N}_L$ in this Hilbert space, which is again infinite \cite{mot}:
\b {\cal N}_L=\sum_{L=0}^\infty\sum_{l=0}^L\sum_{m=-l}^l 1=\infty. \e

\setcounter{equation}{0}
\section{Massive scalar field: plane waves}

Now we consider the massive scalar field in the ambient space formalism. The Laplace-Beltrami operator and the Casimir operator are
proportional:
   $$ \Box_H\equiv-H^2Q^{(1)}_0,$$
where the subscript $0$ in $Q_0$ indicates the action of Casimir operatr on the scalar field. Then the field equation (\ref{sfe}) can be rewritten as:
 \b \label{sfe2} \left(Q^{(1)}_0-  <Q^{(1)}_0>\right)\Phi(x)=0, \;\;\;<Q^{(1)}_0>=m_HH^{-2} +12 \xi=\kappa.\e
For principal series we have $<Q^{(1)}_0>=\frac{9}{4}+\nu^2,\;\; \nu \geq 0$ (\ref{qe1}, \ref{ps}). The parameter $\lambda$ in equation (\ref{lambda}) is $\lambda=i\nu$.
$\Phi$ is a scalar field with
homogeneous degree $\sigma=-\frac{3}{2} \pm i\nu= -\frac{3}{2} \pm \lambda$. The solution of the field equation (\ref{sfe2}) in the ambient space formalism can be written in terms of dS plane waves \cite{brmo}: \b \label{pws}\phi(\xi, x)=\left(\frac{Hx.\xi}{m_H}\right)^\sigma ,\e where $\xi^\alpha$ is a constant five-vector with energy dimensions and it can be chosen as $\xi^2=\eta_{\alpha\beta}\xi^{\alpha}\xi^{\beta}=0.$
These solutions are not globally defined due to the ambiguity of the
phase factor \cite{vikl} and also for $\Re \sigma <0$, which is singular. For obtaining a well defined function one must make use of a proper definition for the $\xi^\alpha$ and $i \epsilon$ prescription \cite{brmo}.  One may consider the
solution in the complex dS space-time $X_H^{(c)}$:
$$ X_H^{(c)}=\{ z=x+iy\in  \C^5;\;\;\eta_{\alpha \beta}z^\alpha z^\beta=(z^0)^2-\vec z.\vec z-(z^4)^2=-H^{-2}\}$$
\b =\{ (x,y)\in  \R^5\times  \R^5;\;\; x^2-y^2=-H^{-2},\; x.y=0\}.\e
Let $T^\pm= \R^5+iV^\pm$ be the forward and backward tubes in $ \C^5$. The domain $V^+$(resp. $V^-)$
stems from the causal structure on $X_H$:
\b V^\pm=\{ x\in \R^5;\;\; x^0\stackrel{>}{<} \sqrt {\parallel \vec x\parallel^2+(x^4)^2} \}.\e
We then introduce their respective intersections with $X_H^{(c)}$,
$$ {\cal T}^\pm=T^\pm\cap X_H^{(c)},$$
which will be called forward and backward tubes of the complex dS space $X_H^{(c)}$. Details can be found in \cite{brmo}. When $z$ varies in ${\cal T}^+$ (or ${\cal T}^-$) and $\xi$ lies in the positive cone ${\cal C}^+$:
\b \xi \in {\cal C}^+=\{ \xi \in \R^5 |\;\; \eta_{\alpha\beta}\xi^\alpha\xi^\beta=0,\; \; \xi^0>0 \},\e
the plane wave solutions (\ref{pws}) are globally defined. When $z \in X_H^{(c)} $ and $\xi \in {\cal C}^+$ the solution is corresponding to the  positive energy in Minkowskian limit \cite{brmo}.

By using the Bros-Fourier transformation \cite{brmo03}, the quantum field operator in this notation is given by:
$$ \Phi(x)=\int_T\left\lbrace\; a(\xi_T,\nu)[(x.\xi_T)_+^{-\frac{3}{2}-i\nu}
        +e^{-i\pi(-\frac{3}{2}-i\nu)}(x.\xi_T)_-^{-\frac{3}{2}-i\nu}]\right.$$
     \b\label{foinam} \left. +a^{\dag}(\xi_T,\nu)[(x.\xi_T)_+^{-\frac{3}{2}+i\nu}
        +e^{i\pi(-\frac{3}{2}+i\nu)}(x.\xi_T)_-^{-\frac{3}{2}+i\nu}]\; \right\rbrace
              d\mu(\xi_T),\e
where $ (x\cdot \xi)_+=\left\{\begin{array}{clcr} 0 & \mbox{for} \;
x\cdot
\xi\leq 0\\ (x\cdot \xi) & \mbox{for} \;x\cdot \xi>0 \\ \end{array}
\right.$ \cite{gesh}.
The annihilation operator $a(\xi,\nu)$ is a homogeneous function with degree $-\frac{3}{2}+i\nu$
      \b \label{aoina} a(\zeta\xi_T,\nu)=\zeta^{-\frac{3}{2}+i\nu}a(\xi_T,\nu).\e
Here $T$ stands for an orbital basis of ${\cal C}^+$. $d\mu_T(\xi)$ is an invariant measure
defined by
$$ d\mu(\xi_T)=i_\Xi w_{{\cal
C}^+}\mid_T,$$ where $i_\Xi w_{{\cal C}^+}$ denotes the $3$-form on
${\cal C}^+$ obtained from the
contraction of the vector field
$\Xi$ with the volume form
$$ w_{{\cal
C}^+}=\frac{d\xi^0\wedge \cdots \wedge d\xi^4 }{d(\xi\cdot \xi)}.$$
Bros et al. have presented two orbital basis \cite{brgamo,brmo}; the hyperbolic-type sub-manifold $T_4=T_4^+
\cup T_4^-$ defined by
\b T^\pm_4=\{ \xi \in {\cal C}^+;\;\; \xi^4=\pm m_H\},\e which is invariant under $SO(1,3)$ group, and the spherical-type submanifold $T_0$ defined by
\b T_0=\{ \xi \in {\cal C}^+;\;\; \xi^0=m_H>0,\;\; \vec{\xi}.\vec \xi+ (\xi^4)^2= m_H^2\},\e which is invariant under $SO(4)$ group. By comparison with the equation (\ref{2orbit}), this orbit is equal to the $\xi^\alpha_u$ where the $\xi^0$ is fixed by $m_H$.

The vacuum state and the "one particle" Hilbert space are depend to the choice of the orbital basis:
  $$  a(\xi_T,\nu)\left|0_T\left>=0, \;\;a^\dag(\xi_T,\nu)\right|0_T\right>=C(\xi_T, \nu)\left|\xi_T;\nu \right>,$$
where $C(\xi, \nu)$ is a normalization constant. It is interesting to note that these two different vacuum states have the same field operator (\ref{foinam}) and the same Wightman two-point function \cite{brmo}. The $SO(1,3)$-orbital basis can be realized by $$\xi_{T_4}^\alpha=\left( k^0,\vec k,\pm m_H\right), \;\; k^0>0, \;\;\ \left(k^0\right)^2-\left(\vec k\right)^2=m_H^2,$$ where $m_H=\nu H^{-1}$. This orbital basis is important in case of considering the null curvature limit \cite{brgamo,brmo}. The $SO(4)$-orbital basis can be chosen as: $$\xi_{T_0}^\alpha=\left( m_H,m_H\vec u, m_H u^4\right)=\xi^\alpha_u, \;\;m_H>0, \;\;\ \left(\vec u\right)^2+\left( u^4\right)^2=1.$$

These two Hilbert spaces ${\cal H}_{T_0}$ and ${\cal H}_{T_4}$, are unitary equivalent \cite{tak,brgamo,brmo}. The identity operators in these two different Hilbert spaces are defined by:
$$ \int_{T_4}d\mu(\xi_{T_4}) \left|\xi_{T_4}; \nu\right>\left< \xi_{T_4};\nu\right| \equiv \1_k , 
 $$
\b \label{identity} \int_{T_0}  d\mu(\xi_{T_0}) |\xi_{T_0}; \nu\left>\right< \xi_{T_0};\nu| 
 \equiv  \1_u . \e
${\cal H}_{T_0}$ and ${\cal H}_{T_4}$ are infinite dimensional Hilbert spaces. Each value of $\xi_{T_0}$ and $\xi_{T_4}$ represents a vector in the respective Hilbert space. The sum or integral over all the possible values of these parameters determine the total number of the quantum states ${\cal N}$. Note that the number of quantum states in these two cases are different. In $SO(1,3)$-orbital basis, it is infinite:
\b  {\cal N}_{T_4}=\int_{T_4} d\mu(\xi_{T_4})=\infty,\e
whereas in the case of $SO(4)$-orbital basis it is finite
\b \label{noft0} {\cal N}_{T_0}=\int_{T_0} d\mu(\xi_{T_0})=  \left(m_H\right)^3 \int_{SO(4)}d\mu(u)=2\pi^2 \left(\nu H^{-1}\right)^3=\mbox{finite value}. \e

Since a maximum length for an observable or an even horizon in dS space exist then a minimum size in the $\xi$-space (or the parameters in Hilbert space) can be defined by using the Heisenberg uncertainty principle. Each point in $\xi$-space represents a vector in Hilbert space and the number of points is infinite mathematically. Since the total volume of $\xi_{T_0}$-space is finite and a minimum length in $\xi_{T_0}$-space exists from uncertainty principle, therefore the total number of points become finite physically. It means that the total number of quantum states in the Hilbert spaces ${\cal H}_{T_0}$ is finite.  
As an example, we consider the compact space $S^1$, where the total volume of space is finite ($2\pi R$) but the total number of points is infinite. If a minimum length such as Planck length $l_p$ exist, the total number of points becomes finite ${\cal N}=2\pi R(l_p)^{-1}$. Therefore the total number of quantum states depends to the scale of energy in the system or to the minimum length.

We would like to emphasize that $\xi^\alpha$ is scale invariant as long as it is defined on the cone:
  \b \label{scalt} \xi'^\alpha=\zeta \xi^\alpha, \;\;\; \mbox{with} \;\; \zeta \neq 0 \Longrightarrow \xi'^\alpha\xi'_\alpha=0=\xi^\alpha \xi_\alpha, \e
which means the scale of $\xi^\alpha$ is arbitrary and the field operator (\ref{foinam}) (and therefore the Wightman two point function) is invariant under this rescaling. Nevertheless, the annihilation and creation operators (\ref{aoina}) (or the Hilbert space) are proportional to this rescaling parameter and the orbital basis as well.  The volume element of the integration also depends on this scale factor:
\b \label{noft02} \int_{T_0} d\mu(\xi_{T_0}')= (\zeta)^3 \int_{T_0} d\mu(\xi_{T_0}),\e
noting that, the identity operators (\ref{identity}) are scale invariant.

What is the physical meaning of these different orbital basis? These different Hilbert spaces are unitary equivalent and therefore the probability amplitude is invariant under the choice of the orbital basis, which is pretty much like the general coordinate transformation on an infinite dimensional Hilbert manifold \cite{lang,chdedi}. The curvature on this Hilbert manifold may be interpreted as an entropic or statistical force similar to the entropical interpretation of the gravitational field \cite{pad,ver,choktu,jac,cawi,lema}. This Hilbert manifold transforms under a new type of scale transformation (\ref{scalt}), since the metric on this manifold (the probability amplitude) is invariant but the states are changed!

What is the physical significance of $\zeta$? Can it be fixed by the theory or by the experimental results? It is important to note that the field operator (\ref{foinam}) is invariant under
the scale transformation (\ref{scalt}) or the value of $\zeta$, where the annihilation operator (\ref{aoina}) depends on the scale $\zeta$. Therefore the Hilbert space and the total number of quantum states ${\cal N}_{T_0}$ are functions of $\zeta$. We already know that ${\cal N}_{T_0}$ is a function of $H$ and $\nu$ (\ref{noft0}), then, $\zeta$ may have a relation with the energy content of dS space-time in the $x^0=$constant or with the size of the spatial part of the space-time. In the next section in order to achieve a better understanding of the ambient space formalism, we shall discuss the governing relation between these two formalisms.

\setcounter{equation}{0}
\section{Plane waves versus partial waves}

In the $SO(4)$-orbital basis $\left(\xi^\alpha=(\xi^0, \xi^0\; {\bf u})\right)$, the relation between the dS plane waves and the partial waves is given by \cite{ta97,gagarota,gasiyo}:
\b \label{plapar}
\left(H x. \xi_{T_0} \right)^{\sigma} =  2\pi^2  \left(\xi^0\right)^{\sigma} \sum_{Llm} \Phi_{Llm}^{\sigma}(X){\cal Y}^{\ast}_{Llm}(u)  ,
\e
where $\sigma=-\frac{3}{2}\pm \lambda= -\frac{3}{2}\pm i\nu$ and
$$
 \Phi_{Llm}^{\sigma} (X) = i^{L-\sigma}\, e^{-i(L+\sigma +3)\rho}(2 \cos{\rho})^{\sigma + 3} \frac{\Gamma(L-\sigma)}{(L+1)!\Gamma(-\sigma)} $$ \b
\phantom{\Phi_{Llm}^{\sigma} (x) =}{}  \times {}_2F_1\big(\sigma + 2, L+\sigma +3; L+2;-e^{- 2i\rho}\big)  {\cal Y}_{Llm}(v) ,\e
where ${}_2F_1$ is the hyper-geometric function. The conformal global coordinates $x^\alpha=(H^{-}\tan \rho, (H\cos\rho)^{-1}{\bf v})$ are being used, where  ${\bf v}=(v^4, \vec v)$ is a quaternion with the norm of $1$. From the linear independence of the hyperspherical harmonics, it is clear that the functions $\Phi_{Llm}^{\sigma} (X)$ are proportional to the mode solution  (\ref{pmc}), $\phi_{Llm}^{\lambda} (X)$ \cite{gasiyo}.

By making use of the Fourier transformation on $S^3$ and the orthogonality of the set of hyper-spherical harmonics we have the integral representation \cite{gasiyo},
\b \label{phix}
\Phi_{Llm}^{\sigma} (X) =  \frac{1}{2\pi^2  (\xi^0)^{\sigma} }  \int_{S^3} d\mu(u)  (Hx\cdot \xi_{T_0} )^{\sigma}  {\cal Y}_{Llm}(u).
\e
Noticing that the functions $\Phi_{Llm}^{\sigma} (X)$ are well def\/ined for all $\sigma$s such as $\Re\sigma < 0$, and so for all scalar de Sitter UIRs \cite{gasiyo}. Defining the following relations:
\b \label{pla2}\left<x|\xi_{T_0};\nu\right>\equiv \left(Hx.\xi_{T_0}\right)^\sigma, \e \b \label{par2} \left<X|Llm;\nu\right>\equiv \Phi_{Llm}^{\sigma} (X) ,\e
and after making use of the identity operator in the respective Hilbert space, one obtains
\b \label{pla3} \left(Hx.\xi_{T_0}\right)^\sigma=\sum_{Llm}  \left<x|L,l,m; \sigma\left>\right< L,l,m;\sigma|\xi_{T_0}, \sigma\right>= \sum_{Llm} \Phi_{Llm}^{\sigma} (X)\left< L,l,m;\sigma|\xi_{T_0}, \sigma\right>.\e
By comparing the equations (\ref{pla3}) and (\ref{plapar}), we can write
$$ \left< Llm;\sigma|\xi_{T_0}, \sigma\right>= 2\pi^2    \big(\xi^0 \big)^{\sigma}   {\cal Y}^{\ast}_{Llm}(u), $$ which are the elements of the unitary matrix of transformation between two Hilbert spaces ${\cal H}_L$ and ${\cal H}_{\xi_{T_0}}$. Again, using the identity operator, the equation (\ref{par2}) can be written in the following form:
$$ \Phi_{Llm}^{\sigma} (X)=  \int_{T_0} d\mu(\xi_{T_0}) \left<x|\xi_{T_0}, \sigma\left>\right< \xi_{T_0},\sigma|Llm; \sigma\right>$$ \b= \int_{T_0} d\mu(\xi_{T_0})  \left(Hx.\xi_{T_0}\right)^\sigma2\pi^2    \big(\xi^0 \big)^{\sigma^{\ast}}  {\cal Y}_{Llm}(u), \e which is the equation (\ref{phix}). It is interesting to note that the Wightman two point function in these two different formalisms is the same \cite{brmo}.

\setcounter{equation}{0}
\section{UIRs of de Sitter group and their Hilbert spaces}

In this section we generalize our previous discussion to the various spin fields (massive or massless) in order to construct the Hilbert spaces and calculate the total number of quantum states. For this purpose we consider the UIRs of dS group and their corresponding Hilbert spaces. Here, we follow the method of Takahashi \cite{tak}. There are different realizations for the construction of the representations and therefore there exist different Hilbert spaces. To gain a better understanding of the problem, we briefly recall this construction for the Poincar\'e group.

The states in the Hilbert space can be described by different notations:
$$ \Psi_{k}\equiv |k>\equiv  f(k).$$  Takahashi have used the notation $f(k)$ \cite{tak} and $\Psi_{k}$ was used by Weinberg  \cite{wei}. In this paper the Dirac notation, $|k>$, is used. The label $k$ denotes all degrees of freedom, which may be continues or discrete. In Minkowski space the energy-momentum four-vector, $k^\mu$, is the best basis choice for the construction of the Hilbert space. The UIR of Poincar\'e group in energy-momentum space can be defined as \cite{wei}
\b \label{poic} {\cal P}^{(M,j)}(\Lambda,a) \left|k^\mu,m_j;j,M\right\rangle=e^{-ia.\Lambda k}\sqrt{\frac{(\Lambda k)^0}{k^0}}\sum_{m_j'}D_{m_j'm_j}^{(j)}(W(\Lambda, k))\left|(\Lambda k)^\mu,m_j';j,M\right\rangle,\e
where $\Lambda \in SO(1,3), \; a^\mu \in \R^4$ and $D_{m_j'm_j}^{(j)}(W(\Lambda, k))$ furnish a certain representation of $SU(2)$ group, which is defined explicitly in \cite{wei}. The two parameters $j$ and $M$ are classifying the UIR of the Poincar\'e group and determining the eigenvalues of the two Casimir operators of the Poincar\'e group. The infinite dimensional Hilbert space ${\cal H}^{(j,M)}_k$ is given by:
$$ \left| k^\mu,m_j;j,M\right\rangle \in {\cal H}^{(j,M)}_k, \;\; (k^0)^2-(\vec k.\vec k)=M^2, \;\; -j\leq m_j\leq j.$$
The total number of the quantum states in this Hilbert space is infinite:
\b \label{nofm} {\cal N}_{{\cal H}^{(j,M)}_k}= (2j+1)\times \int \frac{d^3k}{(2\pi)^3 2\sqrt{\vec k.\vec k+M^2}}=\infty.\e

The energy-momentum four-vector $k^\mu$ does not exist in dS space, but, in the dS ambient space formalism, the tensor and spinor fields, $\Phi(x)$, are a homogeneous function of the coordinates variable $x^\alpha$ (\ref{homog}). It is well known that the homogeneous function has power-law dependence on its arguments \cite{rei}:
$$\Phi(x)\propto \left( x^\beta \right)^\sigma,$$
where $\Phi$ is a homogeneous function of degree $\sigma$.
The only possible dS invariant function for a scalar field in dS ambient space notation, is:
$$\phi(x) \propto \left(x.\xi\right)^\sigma,$$
where $\xi^\alpha$ is a constant five-vector, which transforms under the dS group, {\it i.e.} $\xi .\xi=$constant. This function is singular for $x.\xi=0$ and  $\Re \sigma < 0$, then, it cannot be defined globally in dS space. For complex $\sigma$, $\left(x.\xi\right)^\sigma$ is a many-valued function. In order to turn it into a single-valued well defined function in the whole dS hyperboloid, one must define the proper $\xi^\alpha$ and use a proper $i \epsilon$ prescription, which is explicitly introduced by Bros et al. \cite{brgamo,brmo} and was briefly presented in the previous section. $\xi^\alpha$ is placed on the positive cone by definition: $\xi^\alpha \in C^+=\left\lbrace \xi .\xi=0,\;\;\xi^0>0 \right\rbrace$.

Similar to the energy-momentum four-vector $k^\mu$, the five-vector $\xi^\alpha$ is the best basis choice in this case to construct the Hilbert space. Different choices can be made, by defining the five-vector $\xi^\alpha$, which will result in different Hilbert spaces and also a scale arbitrariness for the definition of $\xi^\alpha$ since it lives on the cone. Four homogeneous spaces were  presented by Takahashi in which cover the principal and discrete series \cite{tak}. Two of these homogeneous spaces are for the principal series, and the other two for the discrete series. For principal series we use the homogeneous space $\xi^\alpha_u=(\xi^0, \xi^0 {\bf u})$, where ${\bf u}$ is a quaternion with norm $|{\bf u}|=1$, and for discrete series we use:
$$\xi^\alpha_B=(\zeta \sinh \kappa, \zeta \cosh \kappa \;\; {\bf q}),\;\;\;|{\bf q}|=|\tanh \kappa |< 1, \;\; 0< \kappa <\infty, $$ 
where their transformation under the dS group are defined in (\ref{sp2tra}) and  (\ref{qt}).

\subsection{Principal series}

The eigenvalue of the Casimir operator $Q^{(1)}_{j,\nu}$ is defined in terms of $j$ and $\nu$ for principal series as:
$$ Q^{(1)}_{j,\nu}=\left[\frac{9}{4}+\nu^2 - j(j+1)\right]I_d, \;\; \; p=\frac{1}{2}+i\nu.$$
The principal series representation of the dS group, which was constructed on a compact homogeneous space $\xi^\alpha_u=(\zeta, \zeta {\bf u})$, is \cite{tak}:
\b U^{(j,\nu)}(g)\left|{\bf u},m_j;j,\nu \right\rangle=|{\bf c}{\bf u}+{\bf d}|^{-2\sigma}\sum_{m_j'}D^{(j)}_{m_jm_j'}\left(\frac{({\bf c}{\bf u}+{\bf d})^{-1}}{|{\bf c}{\bf u}+{\bf d}|} \right) \left|g^{-1}.{\bf u},m_j';j, \nu\right\rangle,  \e
where $\sigma=\frac{3}{2}+i\nu$ and $g^{-1}.{\bf u}= ({\bf a}{\bf u}+{\bf b})({\bf c}{\bf u}+{\bf d})^{-1}$ with $g^{-1}=\left( \begin{array}{clcr} {\bf a} & {\bf b} \\ {\bf c} & {\bf d} \\    \end{array} \right) \in Sp(2,2)$ and the action of the group element on quaternion ${\bf u}$ is defined in (\ref{qt}). $D_{m_jm_j'}^{(j)}$ furnish a certain representation of $SU(2)$ group in a $2j+1$ dimensional Hilbert space $V^j$, which is defined explicitly in \cite{ru70,ru73,vikl}:
$$ D^{(j)}_{mm'}({\bf u})=\left[\frac{(j+m)!(j-m)!}{(j+m')!(j-m')!}\right]^{\frac{1}{2}} \times $$
\b \label{su2} \sum_n \frac{(j+m')!}{n!(j+m'-n)!}\frac{(j-m')!}{(j+m-n)!(n-m-m')!}  u_{11}^{n}u_{12}^{j+m-n}u_{21}^{j+m'-n}u_{22}^{n-m-m'} .\e

The representation $U^{(j,\nu)}(g)$ act on an infinite dimensional Hilbert space ${\cal H}^{(j,\nu)}_{\xi_u}$:
$$ \left|{\bf u},m_j;j,\nu \right\rangle \in {\cal H}^{(j,\nu)}_{\xi_u}, \;\; \xi_u=(\zeta, \zeta {\bf u}), \;\; \zeta >0,\;\; |{\bf u}|=1, \;\; -j\leq m_j\leq j.$$ For the two vectors, $\left|\psi_1\right\rangle , \left|\psi_2 \right\rangle \in {\cal H}^{(j,\nu)}_{\xi_u}$, the scalar product is defined by \cite{tak}:
\b \label{scalarp1} (\psi_1,\psi_2)=\int_{S^3} \left(\psi_1({\bf u}),\psi_2({\bf u})\right)_{V^j} d\mu({\bf u}), \e
where $\left(\psi_1({\bf u}),\psi_2({\bf u})\right)_{V^j}$ is the scalar product in the Hilbert space $V^j$.
One can easily verify that this representation on Hilbert space ${\cal H}^{(j,\nu)}_{\xi_u}$ satisfies \cite{tak}:
$$ U^{(j,\nu)}(g)U^{(j,\nu)}(g')\left|{\bf u},m_j;j,\nu\right\rangle=U^{(j,\nu)}(gg')\left|{\bf u},m_j;j,\nu\right\rangle,$$
\b \label{unitarity} U^{(j,\nu)}(g)\left[ U^{(j,\nu)}(g)\right]^\dag \left|{\bf u},m_j;j,\nu\right\rangle=\left|{\bf u},m_j;j.\nu\right\rangle.\e
After making use the equations (\ref{scalarp1}) and (\ref{unitarity}) and also the references \cite{tak,god,vikl3}, one may choose the states with the following orthogonality condition:
$$    \left\langle{\bf u}',m_{j'};j,\nu \right. \left|{\bf u},m_j;j,\nu \right\rangle =N({\bf u},j,\nu) \delta({\bf u}-{\bf u}') \delta_{m_j m_{j'}}, $$
where $N({\bf u},j,\nu)$ is a normalization constant and $\delta({\bf u}-{\bf u}')$ is the Dirac delta function on $S^3$ \cite{gesh}:
$$ \int_{S^3} \delta({\bf u}-{\bf u}') d\mu({\bf u})=1.$$

Although one has an infinite dimensional Hilbert space  ${\cal H}^{(j,\nu)}_{\xi_u}$, the total number of quantum state is finite in this orbital basis:
\b {\cal N}_{{\cal H}^{(j,\nu)}_{\xi_u}} =(2j+1) \times \int_{T_0} d\mu(\xi_u)=f(H,j,\nu,\zeta)=\mbox{finite value}.\e The total number of quantum states is a function of $H$, $\nu$, $j$ and $\zeta$. This result is due to the compactness of the homogeneous space, which the Hilbert space (or the UIR) is being constructed on it. The existence of this homogeneous space depends on the compactness of the spatial part of the dS space-time. Takahashi has presented another homogeneous space which is not compact and consequently the total number of quantum state on it, is infinite, yet the two spaces are unitary equivalent \cite{tak}. 

The Hilbert space for the quantum field in the dS space-time is an infinite dimensional complex differentiable manifold \cite{lang,chdedi}. Different basis can be chosen for this manifold which are related by the unitary transformation. It is interesting to note that the two-point function (or the probability amplitude) is invariant under the choice of this different basis. The metric on this manifold has the probability interpretation but what is the physical meaning of the curvature on this manifold? We know from the differential geometry that the curvature of the base manifolds is related to the first part of the gravitational field ($g_{\mu\nu}$) and the curvature on the fibre principal is proportional to the other forces propagating on the base manifold, {\it e.g.} gauge fields which contain the conformal gauge gravity (${\cal K}_{\alpha \beta \gamma}$) \cite{ta14}. The curvature on the Hilbert manifold is related to a probability force or a statistical force. This forces have an energy interpretation so we are able to define an entropy. This entropy can be calculated by considering the curvature of the Hilbert manifold, which will be discussed in the forthcoming paper.

\subsection{Discrete series}

The unitary irreducible representation of the dS group for discrete series is constructed on homogeneous space $\xi_B^\alpha=(\zeta\sinh \kappa, \zeta \cosh \kappa\;  {\bf q})$, where the quaternion ${\bf q}$ satisfies the condition $|{\bf q}|=|\tanh \kappa|=r<1$ \cite{tak}:
$$ T^{(j_1,j_2,p)}(g)\left|{\bf q},m_{j_1},m_{j_2};j_1,j_2,p\right\rangle=|{\bf c}{\bf q}+{\bf d}|^{-2p-2} \times $$ \b \label{dseris}\sum_{m_{j_1}'m_{j_2}'}D^{(j_1;j_2)}_{m_{j_1}m_{j_1}';m_{j_2}m_{j_2}'}\left( K(g^{-1},{\bf q})^{-1}\right) \left|({\bf a}{\bf q}+{\bf b})({\bf c}{\bf q}+{\bf d})^{-1},m_{j_1}',m_{j_2}';j_1,j_2,p\right\rangle,  \e
whit $g^{-1}=\left( \begin{array}{clcr} {\bf a} & {\bf b} \\ {\bf c} & {\bf d} \\    \end{array} \right) \in Sp(2,2)$, $j\geq p \geq 1$ and $D_{m_1'm_1;m_2'm_2}^{(j_1,j_2)}$ furnish a certain representation of $SO(4)=SU(2)\times SU(2)$ group, which is defined explicitly in \cite{tak}. The function $K(g^{-1},{\bf q})$ is defined by \cite{tak}:
\b \label{kf} K(g^{-1},{\bf q})=\left( \begin{array}{clcr} ({\bf a}+{\bf b}\bar {\bf q})/|{\bf c}{\bf q}+{\bf d}| &\;\;\;\;\;\;\;\;\;\;\; 0 \\ 0 & ({\bf c}{\bf q}+{\bf d})/|{\bf c}{\bf q}+{\bf d}| \\    \end{array} \right),\e where $\bar {\bf q}=(q^4,-\vec q)$ is a quaternion conjugate and $|{\bf q}|=({\bf q}\bar {\bf q})^{\frac{1}{2}}$ is the norm of quaternion. By using the relation (\ref{kf}), the discrete series representation can be written in the following form:
$$ T^{(j_1,j_2,p)}(g)\left|{\bf q},m_{j_1},m_{j_2};j_1,j_2,p\right\rangle=|{\bf c}{\bf q}+{\bf d}|^{-2p-2} \times $$ \b \label{dseris2}\sum_{m_{j_1}'m_{j_2}'}D^{(j_1)}_{m_{j_1}'m_{j_1}}\left( \frac{({\bf a}+{\bf b}\bar {\bf q})^{-1}}{|{\bf c}{\bf q}+{\bf d}| }\right)
D^{(j_2)}_{m_{j_2}m_{j_2}'}\left(\frac{({\bf c}{\bf q}+{\bf d})^{-1}}{|{\bf c}{\bf q}+{\bf d}|} \right)
 \left|g^{-1}.{\bf q},m_{j_1}',m_{j_2}';j_1,j_2,p\right\rangle,  \e
which is very suitable for the massless field in comparison with the discrete series representation of the conformal group in the next section. It is important to note that $|{\bf a}+{\bf b}\bar {\bf q}|= |{\bf c}{\bf q}+{\bf d}| $ \cite{tak}, then the argument of $D^{(j)}$ is a quaternion with the norm of $1$ , which is a well known representation of $SU(2)$ group (\ref{su2}). 

This representation acts on an infinite dimensional Hilbert space ${\cal H}^{(j_1,j_2,p)}_{\xi_B}$ with the scalar product defined in \cite{tak}.  One can easily verify that this representation on the Hilbert space ${\cal H}^{(j_1,j_2,p)}_{\xi_B}$ satisfies \cite{tak}:
$$ T^{(j_1,j_2,p)}(g)T^{(j_1,j_2,p)}(g')\left|{\bf q},m_{j_1},m_{j_2};j_1,j_2,p\right\rangle=T^{(j_1,j_2,p)}(gg')\left|{\bf q},m_{j_1},m_{j_2};j_1,j_2,p\right\rangle,$$ $$\;\;\;T^{(j_1,j_2,p)}(g)\left[ T^{(j_1,j_2,p)}(g)\right]^\dag\left|{\bf q},m_{j_1},m_{j_2};j_1,j_2,p\right\rangle =\left|{\bf q},m_{j_1},m_{j_2};j_1,j_2,p\right\rangle.$$ Although one has an infinite dimensional Hilbert space  ${\cal H}^{(j_1,j_2,p)}_{\xi_B}$,
$$ \left|{\bf q},m_{j_1},m_{j_2};j_1,j_2,p\right\rangle \in {\cal H}^{(j_1,j_2,p)}_{\xi_B},\;\;\; {\bf q}=r{\bf u} \in \R^4, |{\bf q}|=r<1, \;\; |{\bf u}|=1,\;\; -j\leq m_j\leq j,$$
but the total number of quantum states is finite:
\b {\cal N}_{{\cal H}^{(j_1,j_2,p)}_{\xi_B}}=(2j_1+1)(2j_2+1) \times \int_B d\mu(\xi_B)=\mbox{finite},\e
where \cite{tak}
$$d\mu(\xi_B)=2\pi^2r^3dr d\mu(\xi_u). $$

The representations $T^{j,0,p}$ and $T^{0,j,p}$ are proportional to representations $\Pi^+_{j,p}$ and $\Pi^-_{j,p}$ respectively in the Dixmier notation \cite{dix}. The representations $\Pi^\pm_{j,j}$ in the null curvature limit correspond to the Poincar\'e massless fields \cite{babo,micknied}. They correspond to the two helicitys of the massless fields. Then for massless fields in the de Sitter space we have:
\b T^{0,j;p}(g)\left|{\bf q},m_j;j,p\right\rangle=|{\bf c}{\bf q}+{\bf d}|^{-2p-2}\sum_{m_j'}D^{(j)}_{m_jm_j'}\left(\frac{({\bf c}{\bf q}+{\bf d})^{-1}}{|{\bf c}{\bf q}+{\bf d}|} \right)\left|g^{-1}.{\bf q},m_j';j,p\right\rangle, \e
where  $D_{m_j'm_j}^{(j)}$ furnish a certain representations of $SU(2)$ group (\ref{su2}) \cite{ta14}. These states are defined globally and they are independent of the choice of coordinate system in the dS hyperboloid. For this case the eigenvalue of the Casimir operator $Q_{s,s}^{(1)}$ is: 
$$ Q_{s,s}^{(1)}=2(1+s)(1-s)I_d, \;\;\; j=p=s.$$ The total number of quantum states in this case ${\cal N}_{{\cal H}^{(0,j)}_{\xi_B}}$ is a function of $H$, $j$ and $\zeta$ and it is finite.

We know that in the orbital basis, $T_4$, the $\xi_{T_4}$ is exactly the four-vector of energy-momentum $k^\mu$ in the null curvature limit \cite{brgamo,brmo}. We also demand that every five-vector, $V^\alpha$, in the dS space must be transverse; $x^\alpha V^\beta \eta_{\alpha \beta}=0$ \cite{dir}. In this case $V^\alpha$ transforms as $V'^\alpha=\Lambda^\alpha_\beta V^\beta$ where $\Lambda \in SO(1,4)$. On the contrary, we know that  $x^\alpha \xi^\beta \eta_{\alpha \beta}\neq0$, but, the condition $\xi^\alpha \xi^\beta \eta_{\alpha \beta}=0$ guaranties their transformation under the dS group and $Sp(2,2)$ group (\ref{qt}) and (\ref{2orbit}). We call $\xi^\alpha$ as the generalized energy-momentum. In the global coordinates the spatial sections is a three-sphere of radius $H^{-1}\cosh Ht$. Similar to the above coordinates in the $\xi$-space, we can get
$\xi^\alpha=(\zeta \sinh  \frac{\kappa}{H}, \zeta \cosh  \frac{\kappa}{H}\;{\bf q})$, where ${\bf q}$ is a quaternion with norm $({\bf q}\bar {\bf q})^{\frac{1}{2}}=|\tanh  \frac{\kappa}{H}|$ and $\kappa$ is a parameter with the scale of energy. The scale $\zeta$ in the mathematical framework is completely arbitrary. In the section $VIII$ we will discus the best possible choice for $\zeta$ by imposing the appropriate physical conditions.

\setcounter{equation}{0}
\section{UIRs of the conformal group and their Hilbert spaces}

\subsection{Dirac $6$-cone formalism}

Undesirably, the conformal group acts non-linearly on the Minkowskian coordinates.
Obviating this problem, Dirac proposed a manifestly conformally covariant formulation in
which the Minkowskian coordinates are replaced by some coordinates in
which the conformal group acts linearly. The resultant theory was
then formulated on a 5-dimensional hypercone (named Dirac's
six-cone) in a 6-dimensional space. This method was first used by
Dirac \cite{dirac} to demonstrate the field equations for spinor
and vector fields in $(1+3)$-dimensional space-time in a manifestly
conformal invariant form. This approach to conformal symmetry which leads to best path to exploit the physical symmetry afterwards has been developed by Mack and Salam \cite{masa}.

Dirac's six-cone, or Dirac's projection cone, is defined by \b
y^2\equiv (y^0)^{2}-\vec y^{2}-(y^4)^2+(y^5)^{2}=\eta_{ab} y^a y^b=0 ,\;\;
\eta_{ab}=\mbox{diag}(1,-1,-1,-1,-1,1),\e where $ \;y^{a} \in
\R^{6},$ $a,b=0,1,2,3,4,5$ and  $ \vec y \equiv(y^{1},y^{2},y^{3})$. Reduction to four dimensions is achieved by projection after fixing
the degrees of homogeneity of all the fields. Wave equations,
subsidiary conditions, etc., must be expressed in terms of
operators that are defined intrinsically on the cone. These are
well-defined operators that map tensor fields to tensor fields
with the same rank on the cone $y^2=0$. So, the out coming
equations which are obtained by this method, are conformally
invariant.

The conformal group $SO(2,4)$ acts on the cone as:
$$ y'^a=\Lambda^a_b y^b, \;\;\;\Lambda \in SO(2,4) , \;\; \det \Lambda=1,\;\; \Lambda \eta \Lambda^t= \eta \Longrightarrow.y.y=0=y'.y' $$ If we define a $4\times 4$ matrix, $Y$, as:
$$ Y=\left( \begin{array}{clcr} y^0+iy^5 & \;\;\;\;{\bf p} \\ {\bf \bar p} & y^0-iy^5 \\    \end{array} \right),$$ where ${\bf p}=(y^4, \vec y)$ is a quaternion, then, under a transformation of the conformal group $SU(2,2)$ it transforms as:
$$ Y'=gY\bar g^t, g \in SU(2,2) \Longrightarrow y.y=0=y'.y'.$$
$SU(2,2)$ is the universal covering group of $SO(2,4)$:
 \b SO_0(2,4)  \approx SU(2,2)/ \Z_2 ,\e
which is defined by  \b SU(2,2)=\left\lbrace g=\left(\begin{array}{clcr} {\bf a} &  {\bf b} \\  {\bf c} & {\bf d} \\ \end{array}
        \right),\;\;\mbox{det} g=1 ,\;\;J{\bar g}^t J=g^{-1},
       J=\left(\begin{array}{clcr} \1 &  0 \\  0 & -\1 \\ \end{array}
        \right) \right\rbrace,\e
where ${\bf a},{\bf b},{\bf c}$ and ${\bf d}$ are complex quaternion ${\bf a}={\bf q}_1+i{\bf q}_2$.

The tensor fields $\Psi$ on the Dirac $6$-cone are a homogeneous functions of variable $y^a$ and they are transverse \cite{dirac}:
 $$y^a \frac{\partial}{\partial y^a }\Psi^{cd..}=\sigma \Psi^{cd..},\;\;\;  y_a\Psi^{ab...}=0.$$
In order to project the coordinates on the cone $y^2=0$, to the
$4+1$ dS space we chose the following relations:
\b \left\{
\ba{rcl}
x^{\alpha}&=&(y^5)^{-1}y^\alpha,\\
x^5&=&y^5.\ea\right.\e Note that $x^5$ becomes superfluous when we
deal with the projective cone. By choosing $x^5=H^{-1}$, we obtained exactly the dS hyperboloid and the tensor field which was defined on the Dirac $6$-cone becomes the tensor field on the dS space. For example after doing some algebra, one can show that the following relations are hold for scalar, vector and symmetric rank-$2$ tensor fields on the cone:
$$ \phi(y)=\phi(x),$$
$$ \Psi^a(y)=\left(\Psi^\alpha(y)=K^\alpha(x), \Psi^5(y)=\phi(x)\right),$$
$$ \Psi^{ab}(y)=\left(\Psi^{\alpha\beta}(y)={\cal K}^{\alpha\beta}(x),\Psi^{\alpha 5}(y)= K^\alpha(x), \Psi^{55}(y)=\phi(x)\right),$$
where ${\cal K}^{\alpha\beta}(x)$, $K^\alpha(x)$ and $\phi(x)$ are a symmetric rank-$2$ tensor field, a vector field and a scalar field in dS space respectively. Since these fields are coming from the cone they are conformal invariant. One can also show that they correspond to the discrete series representations of the dS group (next subsection). This can be considered as a dS/CFT correspondence, which is exactly on the dS hyperboloid and exist only for the massless or conformal fields, associated with the discrete series representations, which happens on a four dimensional dS space-time. In the next section we represent the discrete series of the conformal group for a better understanding this correspondence.

\subsection{Discrete UIR of the conformal group}

In the previous works, a variety of realizations of the conformal group UIRs and their corresponding Hilbert spaces have been constructed \cite{masa,mato,ma77,ru72,ru73,anla}. Particularly, Graev's realization of the discrete series is important for our consideration here \cite{ru73}. The homogeneous space is defined by a complex $2 \times 2$ matrix, $Z$, in the domain ${\cal D}$ which is defined by the
constraint of positive definiteness \cite{ru72,ru73}:\b \label{shilov} \1-Z^\dag Z>0, \;\;\;  \1-Z Z^\dag >0 .\e
If we define $Z$ as a quaternion:
$$ Z={\bf q}=\left( \begin{array}{clcr} q^4+iq^3 & iq^1-q^2 \\ iq^1+q^2 & q^4-iq^3 \\    \end{array}, \right) ,$$
then the condition (\ref{shilov}) holds when $|{\bf q}|<1$. It is exactly the homogeneous space ($\xi^\alpha_B$) in which the discrete series representation of the dS group has been constructed on it (\ref{dseris}). The discrete series representation of the conformal group and its corresponding Hilbert space on this homogeneous space, has been explicitly constructed by Ruhl $\big($for value $E_0>j_1+j_2+3$ \cite{ru73} equation (40)$\big)$:
$$ {\cal C}^{(E_0,j_1,j_2)}(g)\left|{\bf q},m_{j_1},m_{j_2};j_1,j_2,E_0\right\rangle=\left[\det({\bf c}{\bf q}+{\bf d})\right]^{-E_0} \times $$ \b \label{dseris2}\sum_{m_{j_1}'m_{j_2}'}D^{(j_1)}_{m_{j_1}m_{j_1}'}\left( {\bf a}^\dag+{\bf q}{\bf b}^\dag\right)D^{(j_2)}_{m_{j_2}'m_{j_2}}\left( {\bf c}{\bf q}+{\bf d}\right) \left|g^{-1}.{\bf q},m_{j_1}',m_{j_2}';j_1,j_2,E_0\right\rangle,  \e
where $g^{-1}=\left( \begin{array}{clcr} {\bf a} & {\bf b} \\ {\bf c} & {\bf d} \\    \end{array} \right) \in SU(2,2)$ and $ g^{-1}.{\bf q}=({\bf a}{\bf q}+{\bf b})({\bf c}{\bf q}+{\bf d})^{-1}.$ $D^{(j_1)}$ and $D^{(j_2)}$ furnish a certain representation of $SU(2)$ group (\ref{su2}) \cite{ru70,ru72,ru73}. The UIR ${\cal C}^{(E_0,j_1,j_2)}(g)$ acts on an infinite dimensional Hilbert space ${\cal H}^{(E_0,j_1,j_2)}_{\xi_B}$. The scalar product was defined in \cite{ru73}.  One can easily verify that this representation on the Hilbert space ${\cal H}^{(E_0,j_1,j_2)}_{\xi_B}$ satisfies \cite{ru72,ru73}:
$$ {\cal C}^{(E_0,j_1,j_2)}(g){\cal C}^{(E_0,j_1,j_2)}(g')\left|{\bf q},m_{j_1},m_{j_2};j_1,j_2,E_0\right\rangle={\cal C}^{(E_0,j_1,j_2)}(gg')\left|{\bf q},m_{j_1},m_{j_2};j_1,j_2,E_0\right\rangle,$$
$${\cal C}^{(E_0,j_1,j_2)}(g)\left[ {\cal C}^{(E_0,j_1,j_2)}(g)\right]^\dag\left|{\bf q},m_{j_1},m_{j_2};j_1,j_2,E_0\right\rangle =\left|{\bf q},m_{j_1},m_{j_2};j_1,j_2,E_0\right\rangle.$$
Although one has an infinite dimensional Hilbert space  ${\cal H}^{(E_0,j_1,j_2)}_{\xi_B}$,
$$ \left|{\bf q},m_{j_1},m_{j_2};j_1,j_2,E_0\right\rangle \in {\cal H}^{(E_0,j_1,j_2)}_{\xi_B},\;\;\; {\bf q} \in \R^4, |{\bf q}|=r<1, \;\; -j\leq m_j\leq j,$$
the total number of quantum state is finite. All possible admissible values for 
$E_0$, $j_1$ and $j_2$ which correspond to the UIR of the conformal group were discussed by Mack \cite{ma77}, but here, our interest lies only on the values:
$ j_1j_2=0, \;\; E_0=j_1+j_2+1,$ with helicity $ j_1-j_2$, where $2j_1$ and $2j_2$ are non-negative integers. These values are in association with 
the massless fields \cite{ma77,ta14}.
 
In the Minkowski space, the massless field equations are conformally
invariant and for every massless representation of the Poincar\'e group
there exists only one corresponding representation in the conformal
group \cite{babo,anla}. In the dS space, for massless
fields, only two representations in discrete series
$(\Pi^{\pm}_{s,s})$ have Minkowskian interpretations. The signs
$\pm$ correspond to the two types of helicity for the massless
field. The representation $\Pi^+_{s,s}$ has a unique extension to a
direct sum of two UIR's ${\cal C}^{(s+1,s,0)}$ and ${\cal C}^{(-s-1,s,0)}$ of
the conformal group $SO_0(2,4)$. Note that  ${\cal C}^{(s+1,s,0)}$ and
${\cal C}^{(-s-1,s,0)}$ correspond to positive and negative energy
representations in the conformal group respectively \cite{babo,anla}.
The concept of energy cannot be defined in the dS space. The UIR of the dS group restricts to the sum of the massless Poincar\'e UIR's ${\cal P}^{(0,s)}_+$
and ${\cal P}^{(0,s)}_-$ with positive and negative energies respectively (\ref{poic}). The following diagrams illustrate these connections
$$ \left. \begin{array}{ccccccc}
     &             & {\cal C}^{(s+1,s,0)}
& &{\cal C}^{(s+1,s,0)}   &\hookleftarrow &{\cal P}^{(0,s)}_+\\
 \Pi^+_{s,s} &\hookrightarrow  & \oplus
&\stackrel{H=0}{\longrightarrow} & \oplus  & &\oplus  \\
     &             & {\cal C}^{(-s-1,s,0)}&
& {\cal C}^{(-s-1,s,0)}  &\hookleftarrow &{\cal P}^{(0,s)}_-,\\
    \end{array} \right. $$        
$$ \left. \begin{array}{ccccccc}
     &             & {\cal C}^{(s+1,0,s)}
& &{\cal C}^{(s+1,o,s)}   &\hookleftarrow &{\cal P}^{(0,-s)}_+\\
 \Pi^-_{s,s} &\hookrightarrow  & \oplus
&\stackrel{H=0}{\longrightarrow} & \oplus  & &\oplus  \\
     &             & {\cal C}^{(-s-1,0,s)}&
& {\cal C}^{(-s-1,0,s)}  &\hookleftarrow &{\cal P}^{(0,-s)}_-,\\
    \end{array} \right. $$    
where the arrows $\hookrightarrow $ designate unique extension. $
{\cal P}^{(0,-s)}_{\pm}$ are the massless Poincar\'e UIRs
with positive and negative energies and negative helicity. 

\setcounter{equation}{0}
\section{Entropy in de Sitter space-time}

In order to calculate the entropy in the dS space, firstly the physical system must be properly defined. In quantum field theory the building block of the universe are the various spin fields including the gravitational fields. The gravitational field can be divided into the two parts. The first part is the dS metrics $g^{dS}_{\mu\nu}$ with the topology $\R \times S^3$ and the kinematical group $SO(1,4)$, which considered as the space-time and behave classically.
The other part is a spin-$2$ conformal field, which propagates on the dS background space time and it is a massless conformal quantum field \cite{ta14}. From the group theoretical point of view it is proved that the second part cannot be described by a rank-$2$ symmetric tensor field in the ambient space notation or Dirac $6$-cone formalism \cite{bifrhe,tatafa,tata}. It must be chosen as a spin-$2$ rank-$3$ mixed symmetric tensor field ${\cal K}_{\alpha \beta \gamma}$, which simultaneously transforms under the dS and conformal groups.

As mentioned, one can define the massive and massless various spin quantum fields in dS space. The massive fields propagate inside the light-cone and are associated to the principal series representation of the dS group, where as the massless fields propagate on the dS light-cone. One of the interesting properties of the massless fields is that they are conformal invariant as long as there is no energy scale in the theory so its field equations are supposed to be scale invariant. The other commune property of massless field theory for $s\geq 1$ is the gauge invariance. The massless field operator transforms as an indecomposable representation of the dS group, and their physical modes are associated to the discrete series representation of the dS and the conformal groups \cite{ta14}. Therefore our system is:
\begin{itemize}
\item{A: The dS background metric, $g_{\mu\nu}^{dS}(x)$, which its effect appears in the other fields as the construction of the Hilbert space. The compact character of the spatial part $(S^3)$ causes to obtain a finite value for the total number of quantum states for the various spin fields. The dS background metrics is considered as a classical field.}
\item{B: Massive fields, with their quantum field operators transform as the UIR of the principal series representation of the dS group.}
\item{C: Massless field, that includes the second part of the gravitational field $({\cal K}_{\alpha\beta\gamma})$.  The physical states of quantum massless field operator simultaneously transforms under the UIR of the discrete series of  dS group and the conformal group.}
\end{itemize}

For simplicity we consider only the interaction between the various fields (parts B and C) with the dS background metrics $g_{\mu\nu}^{dS}(x)$ (part A) and ignore all other interactions. Then the dS universe can be described by a micro-canonical ensemble. In this case the density matrix of the universe is:
\b \rho_{mn}=\rho_n \delta_{mn}=\left\{ \ba{rcl}
\frac{1}{{\cal N}}&,& \mbox{for each of the accessible states}\\ 0&,&  \mbox{for other states }
\ea\right.\e
where $ {\cal N}$ is the total number of quantum states of the dS universe, which is filled by the massive (principal series) and massless (discrete series) various spin fields:
\b {\cal N}=\sum_n \left[c_n^1(j,\nu) {\cal N}_{{\cal H}^{(j,\nu)}_{\xi_{u}}} +c_n^2(j)  {\cal N}_{{\cal H}^{(0,j)}_{\xi_B}}\right].\e 
$n$ is the particle types number, which includes
a specification of its mass, spin, charge, flavour and color. The summation is over the total number of particles exist in the universe (leptons, quarks, Higgs, gauge fields, ...). Based on the standard model of particle physics, $c^1_n$ and $c^2_n$ are finite then the entropy for this micro-canonical ensemble is also finite:
\b S=k_B \ln {\cal N}.\e

For example we consider the universe which is filled by the massive scalar field with definite $\nu$ or $m_H$, then $n=1$ and we have $c_1^1=1$ and $c_1^2=0$. By using the equations (\ref{noft0}) and (\ref{noft02}) and in specific normalization, the total number of quantum states in this case is:
\b {\cal N}={\cal N}_{{\cal H}^{(0,\nu)}_{\xi_{u}}} =a\zeta^3\left(\nu H^{-1}\right)^3,\e where $a$ is a normalization constant and $\zeta$ is an arbitrary function. Therefore the entropy is
\b  S=\ln a+3k_B
\ln \left(  \nu \zeta H^{-1}\right)\equiv 3k_B
\ln \left(  \nu \zeta H^{-1}\right),\e
where the first term can be drooped since entropy is always defined apart from an unknown additive constant.
This is a continuous function of the Hubble parameter $H$ or the area of the dS horizon $A=4\pi H^{-2}$. We fix the function $\zeta$ by the physical conditions. For simplicity we define a new dimensionless function $f(\nu,H)=\left(\nu H^{-1}\;\zeta\right)^3$ where the entropy becomes ($k_B=1$):
$$ S= \ln f. $$

Entropy is a measure of the number of specific ways in which a system may be arranged and therefore it reaches to its maximum in the Minkowski space. The entropy of an isolated system never decreases, because isolated systems spontaneously evolve towards thermodynamic equilibrium, which is the state of maximum entropy.
Entropy is a thermodynamic quantity that helps to keep the account of the energy flow through a thermodynamic process. We know that in the null curvature limit $(H\rightarrow 0)$ the number of quantum states becomes huge (\ref{nofm}) and in the ultimate curvature limit $(H\rightarrow\infty)$ it must turn to the minimum value of the entropy or ${\cal N}$. So we impose following first condition over $f$:
\b \label{condi}\lim_{H\rightarrow0}   f(\nu,H)\longrightarrow \mbox{maximum value},\;\;\; \lim_{H\rightarrow \infty}  f(\nu,H) \longrightarrow \mbox{minimum value}.\e
We can also impose a second condition, using the entropy bounds or the Holographic principle \cite{su,th,suli}:
\b \label{condi2} S_{\mbox{max}} \leq \frac{A}{4G}=\frac{\pi}{GH^2} , \; \Longrightarrow\; \ln f \leq \frac{\pi}{GH^2}.\e
Despite these conditions, there are different possibility for defining this function. We consider some examples here:
\begin{itemize}
\item
For the first choice using the condition (\ref{condi2}) and considering the maximum value for the entropy, we obtain
$$ \ln f=\frac{\pi}{GH^2}\; \Longrightarrow\; f= e^{\frac{\pi}{GH^2}}.$$
The total energy in the dS space for this entropy is deduced from the following relation:
\b \label{e=tds} dE=TdS= \frac{H}{2\pi }dS,\e
consequently, the total energy in the de Sitter space is:
\b \label{energy2} E=\frac{1}{GH}=\frac{1}{2\pi G T}.\e
In the null curvature limit, we obtain $ E=\infty.$ What does it mean? The energy in Minkowski space is bigger than the de Sitter space! Where did we go wrong? We now consider yet another example with a better behaviour in the null curvature limit.
\item
A simple interesting example, which satisfies the condition (\ref{condi}) is $f = 1+\frac{\mu^2}{H^2}$, where $\mu$ is a constant with the energy dimension. This function also satisfies the condition (\ref{condi2}). For this function the entropy is:
\b \label{entro1} S=\ln\left( 1+\frac{\mu^2}{H^2} \right). \e
In the limit $H> \mu$, we have:
$$ S=\sum_{n=1} (-1)^{n-1}\frac{1}{n}\left(\frac{\mu^2}{H^2}\right)^n ,  $$
and, in the limit $H\gg \mu$ the first term remains and the entropy can be written in the following form:
\b\lim_{H\gg \mu} S=\frac{\mu^2}{H^2},\e
which is proportional to the Bekenstein-Hawking entropy or the entropy of the dS space. The total energy in dS space for this entropy (\ref{entro1}) is obtained from the following relation 
$$ E=-\frac{\mu}{\pi} \arctan \left( \frac{H}{\mu}\right)=-\frac{\mu}{\pi} \arctan \left( \frac{2\pi T}{\mu}\right).$$
What is the physical meaning of this negative energy? Gibbons and Hawking have suggested to use the following relation \cite{giha,spstvo}:
\b \label{e=-tds} d(-E)=TdS,\e
instead of the equation (\ref{e=tds}), and then we obtain:
\b E=\frac{\mu}{\pi} \arctan \left( \frac{H}{\mu}\right)=\frac{\mu}{\pi} \arctan \left( \frac{2\pi T}{\mu}\right).\e
In the null curvature limit we have $(H=0=T)$,
\b \label{energy1} E=\frac{\mu}{\pi} \sum_n^\infty \frac{(-1)^n}{2n+1}\left( \frac{2\pi T}{\mu}\right)^{2n+1}=2T-\frac{\mu}{3}\left( \frac{2\pi T}{\mu}\right)^3+...\;\;.\e
In this limit, we obtain $ E=0$, which is a logical result! 
\end{itemize}

There exist the various choice for the function $f$ which satisfy the conditions (\ref{condi}) and (\ref{condi2}), so the remaining questions are; what is the best choice for $f$? What is the physical meaning of this energy $E$? $E$ is an entropic energy and it has an statistical interpretation. In the vacuum all of the quantum fields fluctuate as:
$$ < 0| \Phi^i_{\alpha_1,...\alpha_n}( x) |0>=0,\;\;\; <0| \left[\Phi^i_{\alpha_1,...\alpha_n}( x) \right]^2|0> \neq 0.$$
The effect of the classical background gravitational field $g_{\mu\nu}^{dS}(x)$ on this quantum fluctuation appears as the creation of the some of these fields and it must preserve the other conservation laws of physics. This is similar to the Casimir effect where the role of the parallel plate is being played here by the classical background gravitational field $g_{\mu\nu}^{dS}(x)$. Then, in the absence of this classical background gravitational field (in the Minkowski space), the energy must be zero. Therefore yet another constraint can be defined on the $f$, according to the following relation
\b \label{condi3} \lim_{H\rightarrow 0 }E= \lim_{H\rightarrow 0}\frac{1}{2\pi} \int \frac{Hf'}{f}dH \Longrightarrow 0,\e where $f'=\frac{df}{dH}$.
Can this energy be described as a part of the dark energy?  This question will be considered in our forthcoming paper. Nevertheless and just for the sake of presenting a proper respond to this question, it is obvious that we must consider all of the fields in the dS universe, including the massless fields which will hold an importance role in the answer, provided that the conformal gauge quantum gravity lies in this part. The effect of the interaction between the various spin fields must be added to our calculation which it impose the additional conditions over the function $f$. The interaction is reformulated in ambient space formalism by using the gauge principle \cite{ta14}.

\section{Conclusion and outlook}

The quantum states in the Minkowski space with the symmetrical Poincar\'e group is labelled by $\left |k^\mu,m_j; j,M \right\rangle $, where the eigenvalues of the Casimir operators of the Poincar\'e group are given by $j$ and $M$. Each value of $k^\mu$ and $ m_j$ defines a vector or a state in the Hilbert space. This is the range of $k^\mu$ causing the total number of quantum states to become infinite.

In this paper, the construction of the Hilbert space in the ambient space notation have been discussed. The existence of a specific realization for construction of the infinite dimensional Hilbert space with finite total number of quantum states have been presented. The states are labelled by $\left |\xi^\alpha,m_j; j,\nu \right\rangle $. In this case, for the homogeneous spaces $\xi_u^\alpha$ and $\xi_B^\alpha$, the total numbers of quantum states of our universe are finite.  This property appears due to the compactness of the spatial part of the dS space-time, which means in each slice of $t=$constant, space is unbounded and there is no spatial infinity. Then, the entropy of the dS universe in presence of a massive scalar field has been calculated. We obtain the total number of quantum states or the entropy as an finite arbitrary function $f$, which must to fix. It is logical to assume that the number of the quantum states not only must be a function of $H$ but it also must depend on the size of the spatial part of the universe ($H^{-1} \cosh Ht$). How is this possible? Can we impose this as a condition for fixing the function $f$? Here we have presented three conditions (\ref{condi}), (\ref{condi2}) and (\ref{condi3}) for fixing $f$. In the forthcoming paper, these questions will be studied. 

This method can be generalized to other massive fields (principal series representation of the dS group), and massless fields (discrete series representation), including the conformal gauge gravity as a part of the latter type. By considering the gauge conformal gravity as a principal fibre $[SO(2,4)]$ on the dS base manifold $[SO(1,4)]$. This field is a rank-$3$ mixed symmetry tensor field ${\cal K}_{\alpha\beta\gamma}$ in the dS ambient space formalism \cite{ta14}.

\vspace{1.0cm} \noindent {\bf{Acknowledgements}}: I thank my wife, Sara, who is the source of my encouragement for this work. I would like to thank J.P. Gazeau, J. Iliopoulos, M.R. Masomi Nia, J. Renaud, S. Rouhani and M.R. Tanhayi for their helpful discussions, over some parts of this subject, throughout the years. I hereby dedicate this work to my cousin, Babak Seyed-Hosseini, who is not between us any more.


\end{document}